\documentstyle{elsart}

\begin{document}
\begin{frontmatter}
\title{Ordering Phenomena on Growing Films}

\author{Mehran Kardar}
\address{Physics Department, MIT, Cambridge, MA 02139, USA}
\thanks[NSF]{Supported by NSF Grant No.~DMR-93-03667.}
\begin{abstract}
In many growth processes particles are highly mobile in an active layer
at the surface, but are relatively immobile once incorporated in the bulk.
We study models in which atoms are allowed to interact, equilibrate, and
order on the surface, but are frozen in the bulk.
Order parameter correlations in the resulting bulk material are highly
anisotropic, reflecting its growth history.
In a flat (layer by layer) growth mode, correlations perpendicular to the
growth direction are similar to a two dimensional system in equilibrium,
while parallel correlations reflect the dynamics of such a system.
When the growing film is rough, various couplings between height and
order parameter fluctuations are possible.
Such couplings modify the dynamic scaling properties of surface roughness,
and may also change the critical behavior of the order parameter.
Even the deterministic growth of the surface profile can result in interesting
textures for the order parameter.  

\end{abstract}
\end{frontmatter}

\section{Introduction}

For many technological applications, high quality films are grown by the
process of vapor deposition.
The properties of such films can be quite different from the same material
produced in bulk equilibrium \cite{fro91,roo95}, reflecting their preparation history.
For example, during the growth of some binary alloys, the deposited atoms
are highly mobile on the surface, but relatively immobile in the bulk\cite{roo95}. 
Consequently, the surface fluctuations occurring during the growth process 
are frozen into the bulk. A characteristic feature of such (metastable) phases 
is {\it anisotropic} correlations  related to the growth direction 
which are absent in bulk equilibrium.

A number of models for composite film growth have been introduced in
the past\cite{wel73,ent77,kim85,dav87,YBY90,roo93,kot97}.
Generally in these models, the probability that an incoming atom sticks to a
given surface site depends on the state of neighboring sites in the layer below.
Once a site is occupied, its state does not change any more, and thus the
surface configuration becomes frozen in the bulk.
Such growth rules are equivalent to (stochastic) cellular automata, where
each site is updated in parallel as a function of the states of its neighbors.
Subsequent states of the cellular automaton correspond to successive
layers in the crystal.

It is in general not possible to calculate exact correlation functions for
such (non-equilibrium) growth processes.
The exception occurs in special cases where the growth rules satisfy
a detailed balance condition, relating their stationary behavior to an
equilibrium system of one lower dimension\cite{gri85}.
However, it can be shown that if $d$--dimensional probabilistic
cellular automata with two states, and up-down symmetry, undergo a
symmetry breaking, their critical behavior is identical to the corresponding
Ising model in equilibrium \cite{gri85}. Correlations in time are then
equivalent to those generated by Glauber dynamics of the Ising system.
$(d+1)$--dimensional crystals grown according to the rules of these cellular
automata therefore have an order-disorder phase transition with correlations
perpendicular to the growth direction characterized by the critical exponent
$\nu$, and those parallel to the growth direction by the exponent $\nu z$ of
the $d$--dimensional Ising model ($z$ being the appropriate dynamical
critical exponent).

In Sec.~\ref{model}, I introduce a model for  {\em layer by layer} growth of 
binary films. The atoms on the top layer are assumed to equilibrate completely 
(by surface diffusion {\em or} desorption--resorption mechanisms) 
before another layer is added\cite{dro97}.
Such an assumption is realistic only if the growth rate is much slower
than characteristic equilibration times of the surface layer.  
The model satisfies detailed balance, and can therefore
be analyzed with methods from equilibrium statistical physics.
This discrete model is then used to justify a continuum formulation
to the problem which is identical to the time dependent Landau--Ginzburg
equation for model A dynamics\cite{hoh77}.

In general, a layer by layer growth mode is unstable, and
the role of surface roughness is explored in Sec.~\ref{rough}.
The Kardar, Parisi, Zhang (KPZ) equation describes the
dynamic fluctuations in the  height of an amorphous surface\cite{kpz}.
The interplay between roughness and ordering phenomena is then
considered by introducing simple equations that couple fluctuations
in height and the order parameter.
Long range correlations occur at the critical point for the onset of 
ordering in the surface binary mixture. 
This in turn leads to greater roughness fluctuations,
whose scaling can be explored perturbatively around $d=4$ dimensions\cite{dro98}.

While the order parameter for binary deposition is a scalar, we can more generally
examine the case of a continuous order parameter formed on the surface layer.
There are soft modes associated with  such continuous symmetry breaking
whose coupling to height fluctuations are explored in Sec.~\ref{continu}.
In particular, the deterministic relaxation of the order parameter on an initially
rough surface can in fact be described exactly through a generalized
Cole-Hopf transformation. Interestingly, the relaxation process is super-diffusive
and occurs through coarsening of domains (separated by sharp domain walls)
on surface mounds.

\section{Layer by Layer Growth}
\label{model}

\subsection{Discrete model}

Binary growth is modelled by two kinds of atoms, $A$ and $B$,
which occupy the sites of a $d+1$ dimensional hypercubic lattice. 
Let $\epsilon_{AA}$, $\epsilon_{AB}$, and $\epsilon_{BB}$ denote 
the interaction energies between neighboring atoms of
types $AA$, $AB$, and $BB$ respectively. When
each layer has $N$ sites, there are $2^N$ possible configurations for a layer.
The energy cost for adding a new layer of configuration $\gamma$ on top of one in
configuration $\alpha$ is the sum of the internal energy $E_{\gamma}$ of the
new layer, and the interaction energy $V_{\alpha \gamma}$ with the previous layer. 
These energies are the sums of all local bonds $\epsilon_{ij}$ between
nearest neighbors $ij$ within the new layer, and between
the two layers, respectively. In addition, $E_\gamma$ contains  a chemical
potential $\mu_A N_A + \mu_BN_B$  related to the partial pressures of
$A$ and $B$ atoms in the gas phase.

Assuming that the top layer is in thermal equilibrium, the
conditional probability that it is in configuration $\gamma$, given
configuration $\alpha$ for the layer below, is
\begin{equation}
W_{\gamma \alpha} = {\exp\left[-\beta (E_{\gamma} + V_{\alpha\gamma}) \right]
\over \sum_{\delta} \exp\left[-\beta (E_{\delta} + V_{\alpha\delta})
\right]}\, , \label{W}
\end{equation}
where $T=(k_B\beta)^{-1}$ is the temperature at which the crystal is grown.
After adding many layers, the steady-state probability for finding a configuration
$\gamma$ is determined by the stationarity condition
\begin{equation}
P_{\gamma} = \sum_{\alpha} W_{\gamma\alpha} P_\alpha\,,
\label{stationary}
\end{equation}
which has the solution
\begin{equation}\label{P}
P_\alpha={\sum_\gamma \exp\left[-\beta(E_\alpha + E_\gamma +
V_{\alpha\gamma})\right] \over \sum_{\delta, \nu} \exp\left[-\beta (E_\delta
+ E_\nu + V_{\delta\nu})\right] }\equiv{\sum_\gamma \exp\left[-\beta H_{\alpha\gamma}\right] \over
\sum_{\delta, \nu} \exp\left[-\beta H_{\delta\nu}\right] }\, .
\end{equation}
The above expression is the equilibrium probability for the top layer of a two-layer system, 
obtained after summing over the states of the bottom layer.
Transverse correlation functions (i.e. perpendicular to the growth direction)
are therefore exactly the same as correlation functions in a two-layer system.

From Eqs.~(\ref{W}) and (\ref{P}) it follows that the system
satisfies detailed balance, i.e.,
\begin{equation}
W_{\alpha\gamma}P_\gamma = W_{\gamma\alpha}P_\alpha.
\label{detailedbalance}
\end{equation}
Thus, beyond a transient thickness, the crystal looks the same along
or against the growth direction, and the sequence of layers corresponds to
time evolution of thermodynamic equilibrium states.
This generalizes previous results for cellular automata, which are obtained
by setting the in--plane interactions $E_\alpha$ to zero.
As in such cellular automata, the $(d+1)$--dimensional system has
transverse properties like $d$--dimensional models. In particular, 
phase transitions  occur at the same temperature as
for a $d$--dimensional two-layer system.

Generalizing the model, by allowing several layers at the 
surface to equilibrate, is straightforward. To mimic the large energy of the
impinging particles, as well as their modified environment, we can assign 
each of the top $\ell$ layers from the surface a different temperature, 
through scaled interaction energies depending on its depth. 
The probability that a layer with configuration 
$\gamma$ follows one in configuration $\alpha$ in the bulk is obtained 
by considering the layer at the moment when it is the $\ell$th layer from 
the top, i.e. immediately before its configuration is frozen.
Denoting the configuration of the first $\ell-1$ layers by ${\cal C}_\gamma$ 
and their energy (including the coupling to the $\ell$th layer,
and different interaction constants in the different layers)
by $E({\cal C}_\gamma)$, the conditional probabilities $W_{\gamma\alpha}$
 can be written as
\begin{displaymath}
W_{\gamma \alpha} = {\sum_{{\cal C}_\gamma}\exp\left\{-\beta [E_{\gamma} +
V_{\alpha\gamma} + E({\cal C}_\gamma)] \right\}
\over \sum_{\delta,{\cal C}_\delta} \exp\left\{-\beta [E_{\delta} +
V_{\alpha\delta}+ E({\cal C}_\delta)]
\right\}}\, .
\end{displaymath}
Following the approach for the case $\ell=1$, we can show that the set of weights
\begin{displaymath}
P_\alpha = {\sum_{\gamma,{\cal C}_\gamma,{\cal C}_\alpha} 
\exp\left\{-\beta[E_\alpha + E_\gamma
+ V_{\alpha\gamma} + E({\cal C}_\gamma) + E({\cal C}_\alpha)]\right\}\over
\sum_{\delta, \nu,{\cal C}_\delta,{\cal C}_\nu } \exp\left\{-\beta [E_\delta
+ E_\nu + V_{\delta\nu}+ E({\cal C}_\delta)+ E({\cal C}_\nu)]\right\}},
\end{displaymath}
describe a stationary state. It is easy to verify that this stationary 
solution satisfies detailed balance. 
The  stationary state corresponds to an equilibrium Hamiltonian with $2\ell$ layers, 
with interactions which depend on the distance from the closest surface. 
The top (or the bottom)  layer describes the deposited surface, 
while the middle ($\ell$ or $\ell+1$) layers describe
transverse correlations in the bulk. While the correlations
parallel to the growth direction are more complicated, the
general conclusions  for  $\ell=1$ remain valid.  

\subsection{Continuum formulation}
In the above discrete model, we can use an Ising variable $\sigma_i=\pm 1$ to
indicate if site $i$ is occupied by atom $A$ or $B$.
Close to the critical point, density fluctuations occur over long distances
and universal properties are better captured by considering a coarse-grained
{\rm order parameter\/} $m({\bf x},t)$.
Here ${\bf x}$ labels the $d$ directions transverse to growth, while $t$ which 
indicates time is also proportional to the coordinate parallel to the growth direction.
Hence $m({\bf x},t)$ encodes the time history of the growth process.
From the exact solution of the discrete problem, we know that the behavior
of these configuration is equivalent to the time evolution of a $d$-dimensional
system at equilibrium. 
In the continuum limit, the latter is described by the time-dependent Landau
Ginzburg equation\cite{hoh77}
\begin{equation}\label{LGm}
\partial_t m=K\nabla^2m+rm-um^3+\eta_m({\bf x},t),
\end{equation}
where $\eta_m({\bf x},t)$ is a random noise of zero mean, whose variance is
proportional to the growth temperature.

Away from the critical point at $r=r_c$, fluctuations in $m$ decay over a
transverse correlation length $\xi$, and a longitudinal correlation `time' $\xi^z$.
At the critical point itself, there is no intrinsic scale, and correlations decay as
\begin{equation}\label{mm}
\left\langle m({\bf x},t)m({\bf x'},t') \right\rangle={1\over |{\bf x-x'}|^{-2\chi_m}}
g_m\left( |t-t'|\over |{\bf x-x'}|^z  \right).
\end{equation}
In dimensions $d>4$, criticality occurs for $r=u=0$ (the diffusion equation),
leading to $z=2$ and $\chi_m=(2-d)/2$. On approaching criticality, the correlation
length diverges as $\xi\propto|r-r_c|^{-\nu}$, with $\nu=1/2$.
For $d\leq4$, the nonlinear term $um^3$ is relevant, and the exponents can
be calculated perturbatively\cite{hoh77} in $\varepsilon=4-d$.

\section{Rough Growth}\label{rough}

\subsection{Dynamic roughening}
The layer by layer growth mode cannot be maintained indefinitely, 
and the surface eventually becomes rough\cite{hwa91}.
Let us denote the height of the surface at location ${\bf x}$ at time $t$
by a function $h({\bf x},t)$.
There is considerable evidence from simulations (and some experiments)
that the resulting surfaces exhibit self--affine fluctuations\cite{kar96},
well described by the continuum equation\cite{kpz}
\begin{equation}\label{kpz}
\partial_t h=v_0+\nu\nabla^2h+{\lambda\over2}\left( \nabla h \right)^2+\eta_h({\bf x},t).
\end{equation}
The first three terms in this equation correspond to the average deposition flux, relaxation
by evaporation, and lateral growth, respectively.
The last term represents the fluctuations in the deposition flux, and has zero average.

The self-affine fluctuations in the surface height can be described by dynamic
scaling exponents $\chi_h$ and $z$. For example, the averaged two point 
correlations behave as
\begin{equation}\label{hh}
\left\langle \left[h({\bf x},t)-h({\bf x'},t')\right]^2 \right\rangle= |{\bf x-x'}|^{2\chi_h}
g_h\left( |t-t'|\over |{\bf x-x'}|^z  \right).
\end{equation}
The linear equation for $\lambda=0$ gives diffusive exponents $\chi_h=(2-d)/2$
and $z=2$. Any nonlinearity is relevant in $d\leq2$, while sufficiently large $\lambda$
is required in $d>2$ to produce  a rough phase ($\chi_h\geq0$).

\subsection{Coupling growth and ordering}
There are few studies of the interplay between fluctuations in height and 
the order parameter.
Some numerical simulations have incorporated both elements:
As a model for diamond growth, Capraro and Bar-Yam\cite{cap93} introduced
a variant of ballistic deposition which exhibits sublattice ordering.
Kotrla and Predota\cite{kot97} have examined binary deposition in 1+1
dimensions, resulting in domains with rough surfaces.
In a recent work with Barbara Drossel\cite{dro98}, we took an analytical 
approach to this problem.

The starting point is the continuum Eqs.~(\ref{LGm},\ref{kpz})
describing the order parameter $m({\bf x},t)$, and height  $h({\bf x},t)$, fluctuations.
To these equations we added all terms consistent with the symmetries of
the problem. The lowest order (potentially relevant) terms result in the following
pair of coupled differential equations
\begin{equation}\label{hm}
\left\{ \begin{array}{cl}
\partial_t h =\quad \nu\nabla^2h+{\lambda\over2}\left( \nabla h \right)^2 +\zeta_h\qquad\qquad\qquad
&-{\alpha\over2}m^2\\ 
\partial_t m =K\nabla^2m+rm-um^3+\zeta_m +a\nabla h\cdot\nabla m &+b m\nabla^2h
+{c\over2}m\left( \nabla h \right)^2
\end{array}\right.\, .\quad
\end{equation}
(Note that these equations satisfy the symmetry $m\mapsto -m$.)
Fluctuations of the surface are modified by coupling to the order parameter,
through the term proportional to $\alpha m^2$.
There are also three coupling constants $a$, $b$, and $c$, which modify
the order parameter fluctuations due to coupling to $h$.

As long as the binary mixture is disordered ($r>r_c$), fluctuations in $m$ and hence
$m^2$, are short--ranged, and $\alpha m^2$ acts as another source of white noise.
The surface fluctuations should thus scale with the standard KPZ exponents.
However, the range of correlations increases as $r\to r_c$ and $\xi\sim|r-r_c|^{-\nu}\to\infty$.
This modifies (most likely increases) the overall amplitude of surface roughness,
and height fluctuations over a scale $L$ behave as
\begin{equation}\label{hiT}
\sqrt{\left\langle \delta h^2(L,r) \right\rangle}= \xi^{\chi_h^c-\chi_h}L^{\chi_h}g\left( L/\xi \right),
\end{equation}
where $\chi_h^c$ is the roughness exponent at criticality, which is discussed next.

\subsection{Critical roughness}
Under a change of scale ${\bf x}\mapsto b{\bf x}$, $t\mapsto b^zt$,
$h\mapsto b^{\chi_h}h$, and $m\mapsto b^{\chi_m}m$, the non-linear coefficients
in Eqs.~(\ref{hm}) scales as $x\mapsto b^{y_x}x$, with
\begin{eqnarray}\label{ys}
y_\lambda=y_a=y_b=\chi_h+z-2,\,\,\, y_c=2\chi_h+z-2,\,\,\, y_\alpha=2\chi_m-\chi_h+z .
\end{eqnarray}
The critical point in dimensions $d\geq4$ occurs at $r=u=0$. The linear diffusion
equations at this point result in the bare field dimensions $\chi_h^0=\chi_m^0=(2-d)/2$.
Taking account of the non-linearities, we observe the following behaviors.
\begin{itemize}
\item
$d>6:$ All non-linearities are (perturbatively) irrelevant; the surface is smooth, 
and the order parameter goes through a classical phase transition.
\item
$4<d<6:$  The leading non-linearity is the term $\alpha m^2$ describing the correlated
noise acting on the surface height, with ($z=2$)
\begin{equation}\label{yalpha}
y_\alpha^0=4-d-\chi_h^c=3-d/2.
\end{equation}
In these dimensions, the correlated noise is more relevant than the white noise from
the flux variations\cite{med89}.
The correct result can in fact be obtained simply by setting $y_\alpha$ to zero,
leading to critical height fluctuations with
\begin{equation}
\chi_h^c=4-d >{2-d\over 2}\quad.
\end{equation}
Note that while the roughness exponent is larger than its bare value, it is still negative.
The scaling of the order parameter is not modified, and $\chi_m=(2-d)/2$.
\item
$d\leq4:$ When the roughness exponent is positive, all the couplings
$\lambda$, $a$, $b$, and $c$ become relevant. Also in $d\leq4$, the critical
point of the Landau-Ginzburg model is no longer at $r=u=0$,
and a full renormalization group (RG) study is called for\cite{dro98}.
Ignoring the feedback from height fluctuations to the order parameter,
we find to leading order an RG equation of the form
\begin{equation}
{1\over(\alpha\lambda)}{d(\alpha\lambda)\over d\ell}=
\varepsilon-C(\alpha \lambda),
\end{equation}
where $\varepsilon=4-d$, and $C$ is a {\em positive\/} constant.

There is a fixed point at $\alpha\lambda=\varepsilon/C$, with roughness
exponent $\chi_h^c=0$.
In $d>4$, this is an unstable fixed point governing a transition between
flat ($\chi_h^c=4-d<0$) and rough phases (occurring  for $\alpha\lambda<-(d-4)/C$).
For $d<4$, this fixed point is stable and attracts all points with $\alpha\lambda>0$.
Negative values of $\alpha\lambda$ flow to a rough phase which is not
perturbatively accessible.
Including all non-linearities in the equation for $m$ complicates the analysis,
but we did not find a fixed point whose critical behavior is different
from the ordinary Landau--Ginzburg model (at least to lowest order).
\end{itemize}

\section{Continuous Order}\label{continu}
\subsection{Stochastic evolution}
The situation on the ordered side of the phase transition is more complex.
The analogy to the dynamics of the lower-dimensional system
suggests that the leading process is the gradual coarsening of the
ordered domains. Such domains would then appear as cone-shaped
columns in the bulk film, a reasonably common feature of growth textures.
However, more work is necessary to verify and quantify this picture.

Another interesting situation is when the symmetry breaking involves
a continuous, rather than a discrete (Ising like), order parameter.
For example, we may consider deposition of spins which can realign
on the surface but are frozen in the bulk.
More interestingly, the growth of crystals involves translational and
orientational symmetry breakings in the plane.
In the simplest case of a vector order parameter, we can simply generalize
Eqs.~(\ref{hm}) by replacing the scalar $m$ with an $n$--component
vector ${\vec m}({\bf x},t)$.
While the discussion of critical roughening is not significantly modified
from the Ising case ($n=1$), new issues arise pertaining to the ordered phase.

The most common excitations of the broken symmetry phase are 
{\em soft (Goldstone) modes}, which can in principle couple to the surface roughness.
The simplest example is provide by the XY model ($n=2$), where the
direction of the vector can be described by an angular field $\theta({\bf x},t)$.
Including the lowest order terms which satisfy rotational symmetry
leads to the coupled equations of motion
\begin{equation}\label{htheta}
\left\{ \begin{array}{cl}
\partial_t h = \nu\nabla^2h+{\lambda\over2}\left( \nabla h \right)^2 +\zeta_h\qquad
&-{\alpha\over2}\left( \nabla \theta \right)^2\\ 
\partial_t \theta =\qquad\qquad K\nabla^2\theta+\zeta_\theta\qquad &+a\nabla h\cdot\nabla \theta 
\end{array}\right.\, .\quad
\end{equation}
Interestingly, these are precisely the equations proposed in Refs.~\cite{ert92,ert93}
in the contexts of moving flux lines and drifting polymers.
In these contexts, the above equations have been studied by RG analysis
(around $d=2$), and by numerical simulations (in $d=1$).
In particular, in $d=1$ the KPZ exponents ($\chi_h=1/2$ and $z=3/2$) are
recovered for the surface roughness, while the angular fluctuations remove
any long-range order.
Further analysis is again necessary for the case $d=2$. 
Specifically, an important aspect of the field $\theta$ not present in the earlier
studies is its angular nature.
It could thus include vortices which are {\em topological defects}.
Such defects typically play an important role in equilibrium two dimensional systems,
and have been recently considered in a number of related
non-equilibrium situations\cite{ara98}.

\subsection{Deterministic textures}
It is well-known that the non-linear KPZ equation can be recast as a linear
diffusion equation through the Cole-Hopf transformation.
This transformation can in fact be generalized to describe the coupling
of the surface height to a vector order parameter.
Consider a field of unit spins, $|{\vec s}({\bf x},t)|=1$, and set
\begin{equation}\label{CH}
{\vec W}({\bf x},t)=\exp\left[ \lambda h({\bf x},t)\over 2\nu \right]{\vec s}({\bf x},t).
\end{equation}
A diffusive equation of the field ${\vec W}({\bf x},t)$, as
\begin{equation}\label{vW}
\partial_t {\vec W}=\nu\nabla^2{\vec W}+{\lambda\over 2\nu}\eta_h({\bf x},t){\vec W},
\end{equation}
can be recast into the pair of coupled differential equations
\begin{equation}\label{hs}
\left\{ \begin{array}{cl}
\partial_t h &= \nu\nabla^2h+{\lambda\over2}\left( \nabla h \right)^2 +\left( 2\nu^2\over\lambda \right){\vec s}\cdot\nabla^2{\vec s}+\eta_h({\bf x},t)\nonumber \\ 
\partial_t {\vec s}&=\nu\left[\nabla^2{\vec s}-\left({\vec s}\cdot\nabla^2{\vec s}\,\right)  
\, {\vec s}\,\right]+\lambda\nabla h\cdot\nabla {\vec s} 
\end{array}\right.\, .\quad
\end{equation}
Note that the transverse component of $\nabla^2{\vec s}$ contributes to 
$\partial_t{\vec s}$, thus ensuring that the magnitude of ${\vec s}$ is not changed
in time, while the longitudinal component of this quantity couples to the surface height.

It can be checked easily that for $n=2$, the parametrization 
${\vec s}=\left( \cos\theta,\sin\theta \right)$, reduces Eqs.~(\ref{hs}) to
Eqs.~(\ref{htheta}) in the special limit of $\alpha\lambda=4\nu^2$,
$K=\nu$, $a=\lambda$, and $\eta_\theta=0$.
It is also possible to construct other Cole-Hopf transformations 
for cases when $\alpha\lambda<0$\cite{ert93}.

Starting from any arbitrary initial condition at $t=0$,
the  deterministic limit ($\eta_h=0$) of Eqs.~(\ref{vW}-\ref{hs}) is easily solved
using the diffusion kernel, as
\begin{eqnarray}\label{spW}
{\vec W}({\bf x},t)&=&\exp\left[ \lambda h({\bf x},t)\over 2\nu \right]{\vec s}({\bf x},t)\nonumber\\
&=&\int {d^d{\bf x'}\over \left( 4\pi\nu t \right)^{d/2}}
\exp\left[-{\left( {\bf x-x'} \right)^2\over 4\nu t}+ {\lambda h({\bf x'},0)\over 2\nu}\right]
{\vec s}({\bf x'},0).
\end{eqnarray}
The saddle--point evaluation of the above integral (formally exact as $\nu\to0$)
captures the long time behavior of the solution. 
The surface profile
\begin{equation}\label{sph}
h({\bf x},t)=\min_{{\bf x'}}\left[h({\bf x'},0)-{\left( {\bf x-x'} \right)^2\over 2\lambda t}  \right],
\end{equation}
consists of a set of parabolic mounds centered at locations ${\bf x'}={\bf x_0}({\bf x},t)$
corresponding to high points of the initial surface.
Note that the evolution of the surface profile in this limit is independent of ${\vec s}$. 
The evolution of spins on the other hand is completely controlled by the surface
height, and given by
\begin{equation}\label{spW}
{\vec s}\,({\bf x},t)={\vec s}\left({\bf x_0}({\bf x},t),0\right),
\end{equation}
i.e. each of the surface mounds carries the spin of its initial high point!
Such behavior is quite different from the diffusive evolution of spins in
the absence of coupling to the surface profile.
For self-affine initial surface profiles, the relaxation of the height and spins
is now both diffusive.
Furthermore, the spin textures produced by this process are domains
separated by sharp domain walls, very different from the soft modes and
vortices that characterize diffusive relaxation.
Similar extensions of the Cole-Hopf transformation to matrix order 
parameters are also possible\cite{ert93}, and could for example 
describe relaxation of crystalline substrates.

\begin{ack}
Most of the research described in this paper has been in collaboration
with Barbara Drossel. 
\end{ack}

\end{document}